\documentclass[fleqn,a4,12pt]{article}
\usepackage{amsmath}
\usepackage{tabls}
\usepackage{amsfonts,dsfont,latexsym,stmaryrd,array,graphicx,epsf,pstricks,pst-node,float}

\numberwithin{equation}{section}
\newcommand{\ii}{\mathrm{i}}

\newcommand{\dd}{\mathrm{d}}

\newcommand{\e}{\mathrm{e}}
\newcommand{\ket}[1]{\left|#1\right\rangle}
\newcommand{\bra}[1]{\left\langle #1\right|}

\newcommand{\ft}[2]{{\textstyle\frac{#1}{#2}}}
\def\tilde{\widetilde}
\def\1bar{1\hskip -.275cm -}
\def\2bar{2\hskip -.275cm -}
\def\3bar{3\hskip -.275cm -}

\newsavebox{\uuunit}

\newcommand{\nc}{\newcommand}
\nc{\VEC}{\overrightarrow} \nc{\la}{\lambda} \nc{\alf}{\alpha}
\nc{\tht}{\theta} \nc{\eps}{\epsilon} \nc{\ga}{\gamma}
\nc{\Ga}{\Gamma} \nc{\De}{\Delta} \nc{\de}{\delta}
\nc{\si}{\sigma} \nc{\ka}{\kappa} \nc{\om}{\omega}
\nc{\qq}{\quad\quad} \nc{\nf}{\infty} \nc{\dl}{\mathop{\smash{\cal
L}}} \nc{\ol}{\overline} \nc{\beq}{\begin{equation}}
\nc{\barr}{\begin{array}} \nc{\earr}{\end{array}}
\nc{\eeq}{\end{equation}} \nc{\beqa}{\begin{eqnarray}}
\nc{\dst}{\displaystyle}\nc{\pt}{\partial}
\nc{\eeqa}{\end{eqnarray}} \nc{\nnb}{\nonumber}
\nc{\bs}{\backslash}        \nc{\mbb}{\mathbb}
\nc{\brm}{\begin{remunerate}} \nc{\erm}{\end{remunerate}}
\nc{\vareps}{\varepsilon} \nc{\tb}{\tilde\beta_0} \nc{\ts}{\tilde
s} \nc{\tth}{\tilde \theta}
\nc{\llangle}{\langle\!\langle} \nc{\rrangle}{\rangle\!\rangle}
\newcounter{muni}
\usecounter{muni}
  \nc{\lapdec}{\mathop{\Delta}}

\newenvironment{remunerate}{\begin{list}{{\rm \arabic{muni}.}}
{\usecounter{muni}
\setlength{\leftmargin}{0pt}\setlength{\itemindent}{38pt}}}{\end{list}}
\nc{\cre}{\color[rgb]{1.00,0.00,0.00}}
\nc{\cgr}{\color[rgb]{0.00,1.00,0.00}}

\newcommand{\be}{\begin{equation}} \newcommand{\ee}{\end{equation}}
\newcommand{\bea}{\begin{eqnarray}} \newcommand{\eea}{\end{eqnarray}}
\newcommand{\ben}{\begin{displaymath}}
\newcommand{\een}{\end{displaymath}}

\makeatletter
\def\hlinewd#1{%
\noalign{\ifnum0=`}\fi\hrule \@height #1 %
\futurelet\reserved@a\@xhline} \makeatother

\newcommand{\eq}{\begin{equation}}
\newcommand{\eqx}{\end{equation}}

\begin{document}
\input{epsf}
\topmargin 0pt
\oddsidemargin 5mm
\headheight 0pt
\headsep 0pt
\topskip 9mm
\pagestyle{empty}

\vspace*{110pt}

\begin{center}

{\large \bf {Quasilocality of joining/splitting strings from \\
\hspace*{1.0cm}coherent states }}

\vspace*{26pt}

{\sl P.-Y.\ Casteill}$\,^{a}$, {\sl R.A.\ Janik}$\,^{b}$,
{\sl A.\ Jarosz}$\,^{a}$ and
{\sl C.\ Kristjansen}$\,^{a}$

\vspace{10pt}

$^a$~The Niels Bohr Institute, Copenhagen University\\
Blegdamsvej 17, DK-2100 Copenhagen \O , Denmark.\\
\vspace{10pt}

\vspace{10pt}

$^b$~Institute of Physics, Jagellonian University,\\
Reymonta 4, PL 30-059 Krakow, Poland.\\
\vspace{10pt}

\vspace{20pt}

\end{center}

\begin{abstract}

\noindent
Using the coherent state formalism we calculate matrix elements of the
one-loop non-planar dilatation operator of ${\cal N}=4$ SYM between
operators dual to folded Frolov-Tseytlin strings and observe a curious
scaling behavior. We comment on the {\it qualitative} similarity of
our matrix elements to the interaction vertex of a string field theory.
In addition, we present
a solvable toy model for string splitting and joining.
The scaling behaviour of
the matrix elements suggests that the contribution to the genus one
energy shift coming from semi-classical string splitting and joining
is small.

\end{abstract}

\newpage

\pagestyle{plain}

\setcounter{page}{1}

\section{Introduction}

Integrability has played a key role in recent years exploration of
planar ${\cal N}=4$
SYM~\cite{Minahan:2002ve,Beisert:2003tq,Beisert:2003yb} as well as
non-interacting type IIB string theory on $AdS_5\times
S^5$~\cite{Bena:2003wd,Dolan:2003uh}, tied together by the AdS/CFT
correspondence~\cite{Maldacena:1997re}. Whereas integrability  is
expected to break down beyond the planar/non-interacting limit ---
most clearly demonstrated by the lift of degeneracies of anomalous
dimensions in the gauge theory~\cite{Beisert:2003tq} --- the
AdS/CFT correspondence could still be
valid~\cite{Maldacena:1997re,'tHooft:1974hx}.
Lacking the framework of integrability, tests of the AdS/CFT
correspondence beyond the planar limit have proved difficult. Even
in the BMN limit~\cite{Berenstein:2002jq} where the free string
theory can actually be quantized no conclusive tests exist. For an
up to date review, see~\cite{Grignani:2006en}. The gauge theory
calculations, although described efficiently by a quantum
mechanical Hamiltonian~\cite{Beisert:2002ff}, are plagued by huge
degeneracy problems~\cite{Freedman:2003bh}. The string theory
computations on their side suffer from the existence of several
competing proposals for the three string vertex of light cone
string field theory and from the necessity of truncating the
vertex to a subset of decay channels. Although the BMN limit seems
to be the most tractable one as regards the analysis of the
non-planar sector of the theories it might be instructive to
perform the analysis in other limits as well. A limit which has
been instrumental in the investigation of the
planar/non-interacting case is the Frolov-Tseytlin
limit~\cite{Frolov:2003qc}. A first step in the direction of
extending the analysis of this limit to the non-planar/interacting
situation was taken in~\cite{Peeters:2004pt} where the decay of a
folded Frolov-Tseytlin string~\cite{Frolov:2003xy}
was described using semi-classical
methods. Based on the investigations performed it was argued that
the integrability observed for the free string may survive in
certain decay channels. In the present paper we attack the
non-planar Frolov-Tseytlin limit from the gauge theory side. Using
a coherent state approach we calculate matrix elements of the
one-loop non-planar dilatation generator of ${\cal N}=4$ SYM
between operators dual to folded Frolov-Tseytlin strings rotating
on $S^3\subset S^5 \subset AdS_5\times S^5$.

We begin in section~\ref{dilatation} by presenting the form of the
one-loop non-planar dilatation operator in the $SU(2)$ sector
of ${\cal N}=4$ SYM.
Subsequently, in section~\ref{coherent} we review the coherent state
description of the operator dual to the folded Frolov-Tseytlin string.
Section~\ref{matrixelements} deals with the calculation of matrix elements for
the gauge theory equivalent
 of string joining and string splitting.
In section~\ref{integrable} we describe
 a solvable toy model for the decay of the folded string
which unfortunately is only a very crude approximation to the actual model.
Finally, section~\ref{discussion} contains a discussion.

\section{The one-loop non-planar dilatation operator \label{dilatation}}

\begin{figure}[!h]
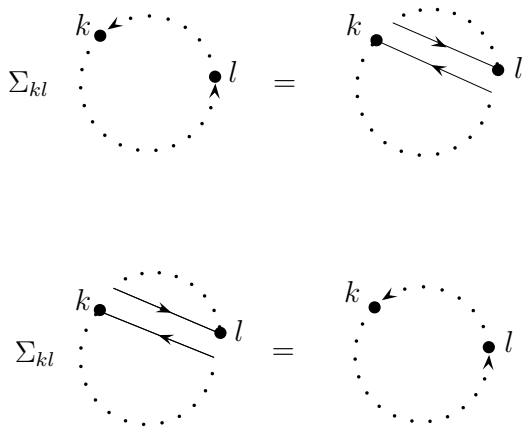

  \centering{\resizebox{6.5cm}{!}{\input{1in2a.pic}}\\
 \resizebox{6.5cm}{!}{\input{2in1a.pic}}}\vspace*{-2cm}
  \caption{Splitting and joining of chains by $\Sigma_{kl}$.}\label{fig:1in2}
\end{figure}

We consider the $SU(2)$ sector of ${\cal N}=4$ SYM consisting of multi-trace
operators built from the two complex scalar fields $Z$ and $\Phi$. In this
sub-sector the complete one-loop dilatation operator can be
expressed as~\cite{Beisert:2002bb,Beisert:2003tq}
\beq
H=-\frac{g_{\mbox{\tiny{YM}}}^2}{8\pi^2} {\mbox {Tr}} [\Phi,Z]
[\check{\Phi},\check{Z}],
\hspace{0.7cm} \check{Z}=\frac{\delta}{\delta Z},
\eeq
or equivalently
~\cite{Bellucci:2004ru,Peeters:2004pt}:
\beq
H=H_{P}+H_{NP},
\eeq
where
\beq
H_P={\lambda}\sum_{k} (1-P_{k,k+1}), \hspace{0.7cm}
\lambda=\frac{g_{\mbox{\tiny{YM}}}^2 N}{8\pi^2},
\label{planar}
\eeq
and
\beq
H_{NP}=\frac{\lambda}{N}\sum_{k,\,l\neq k+1} (1-P_{k,l})\
\Sigma_{k+1,l},
\label{ham1}
\eeq
with $H_P$ being the planar part and $H_{NP}$ the non-planar one.
Here the indices refer to the position of the fields inside the operator
on which $H$ acts. The indices are periodically identified as dictated by the
trace structure of the operator. The operator $P_{k,l}$ simply interchanges
indices $k$ and $l$. Furthermore, if one represents an operator as a
set of fields
plus a permutation element giving the ordering of the fields,
then $\Sigma_{k,l}$ is just the transposition
$\sigma_{k,l}$ applied on this permutation~\cite{Bellucci:2004ru}.
A useful way of describing
the effect of having acted with $\Sigma_{k,l}$ on a chain of fields is
the following (see also fig.~\ref{fig:1in2}) :
\begin{center}
  \emph{The site that was going to k goes to l and vice versa.}
\end{center}

\section{Folded string duals using coherent states \label{coherent}}

\subsection{The Frolov-Tseytlin folded string}
We wish to consider operators dual to the folded Frolov-Tseytlin string spinning
on $S^3\subset S^5\subset AdS_5\times S^5$ with two large angular momenta
$(J_1,J_2)$.
More precisely, we consider the limit $J_1,J_2\rightarrow\infty$ with
$\frac{J_1}{J_2}$ finite. A semi-classical analysis of the string in question
yields that its energy has the following expansion~\cite{Frolov:2003xy}
\beq
E=J\left(1+\frac{\lambda}{J^2}\, {\cal E}_0+
\frac{\lambda^2}{J^4}\,{\cal E}_0^{(1)}+
\ldots\right), \hspace{0.7cm}J=J_1+J_2,
\eeq
with the gauge coupling constant $\lambda$ appearing via the AdS/CFT
dictionary $\frac{R^2}{\alpha'}=\sqrt{\lambda}$~\cite{Maldacena:1997re}
and where we also assume that $\frac{\lambda}{J^2}$ is finite.
The term of linear order in $\lambda$ is found to be
\beq
{\cal E}_0=16\,K(m)\left(E(m)-(1-m)K(m)\right),
\label{E1}
\eeq
where $K(m)$ and $E(m)$ are the complete elliptic integrals of the
first and the second kind respectively.\footnote{Here and in
the following we use the Mathematica definition of elliptic functions and
integrals.}
The parameter $m$ is determined by
\beq
\frac{J_2}{J}=1-\frac{E(m)}{K(m)}.
\label{J2overJ}
\eeq

\begin{figure}[h]
\centerline{\hbox{
   \epsfxsize=6.0in
   \epsffile{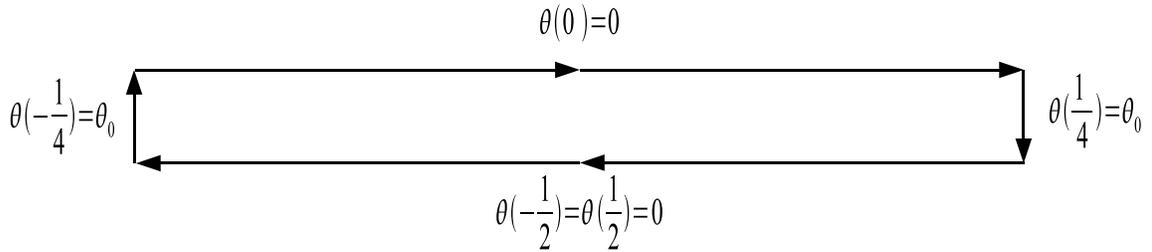}}
  }
  \caption{Different values of  $\theta=\theta(\sigma)$ along the string.}
\label{fig:thetaidentifications}
\end{figure}

The gauge theory dual of the folded Frolov-Tseytlin string is a complicated
linear combination of single trace operators each containing
$J_1$ $\Phi$'s and $J_2$ $Z$'s~\cite{Frolov:2003xy,Beisert:2003ea}.
It is characterized by being an eigenstate of the one-loop planar dilatation
operator, $H_P$, cf.\ eqn.~(\ref{planar}), with eigenvalue given by
$\frac{\lambda}{J}{\cal E}_0$. A more efficient way of describing the dual
is by means of $SU(2)$ spin-$1/2$ coherent states.
To introduce these, let
us denote the two normalized eigenstates of $S_z$ by $\ket{\uparrow}$
and $\ket{\downarrow}$. These states have the inner product
\bea
\bra{\uparrow}\uparrow\rangle&=&\bra{\downarrow}\downarrow\rangle=1,
\nonumber\\
\bra{\uparrow}\downarrow\rangle&=&\bra{\downarrow}\uparrow\rangle=0.
\nonumber
\eea
The relevant coherent states then take the form
\beq
\ket{\vec{n}}=\cos\theta\,\ket{\uparrow}+\e^{-\ii\,\varphi}\sin\theta\,
\ket{\downarrow},
\eeq
where the angles $\theta\in[0,\frac{\pi}{2}]$ and $\varphi\in[0,2\pi]$
parametrize
a unit three vector $\vec{n}$ by
\beq
\vec{n}=\left(\cos2\theta \sin\varphi,\sin2\theta \sin\varphi,
\cos\varphi
\right).
\eeq
The folded string dual can now be described as a state of a $SU(2)$ spin
chain of length $J$ having a coherent state vector at each
site~\cite{Kruczenski:2003gt}. Without
loss of generality we will take $J$ to be a multiple of four, in order
for the spin chain to reflect as closely as possible the symmetries of the
folded string (the entire string profile follows from its definition on
a quarter period). The state representing the string
thus reads
\beq
\ket{\bf{n}}=\ket{\vec{n}_{-\ft J2}}\otimes\ket{\vec{n}_{-\ft J2+1}}\cdots\otimes\ket{\vec{n}_{\ft J2}},
\eeq
where obviously
\beq
\ket{\vec{n}_k}=\cos\theta_k\,\ket{\uparrow}+
\e^{-\ii\,\varphi_k}\sin\theta_k\,\ket{\downarrow}.
\eeq
Here the planar energy of the string is obtained as
$\frac{\lambda}{J} {\cal E}_0=\langle {\bf n}|H_p|{\bf n}\rangle$.
In the long wavelength limit where $\theta_k$ and $\varphi_k$ vary only
slowly and where $J\rightarrow \infty$, which exactly corresponds to
the Frolov-Tseytlin limit,
one can replace the $\theta_k$ and $\varphi_k$ by continuous functions
$\theta_k\rightarrow \theta(\sigma=\frac{k}{J})$ and
$\varphi_k\rightarrow \varphi(\sigma=\frac{k}{J})$ and one can derive an
effective sigma model action describing the model. The cyclicity
property of the gauge theory operator translates into the
requirement of vanishing momenta in the $\sigma$ direction, which reads
\beq
{\cal P}_\sigma=-\frac{1}{2}
\int_{-\ft12}^{\ft12}\cos(2\theta) \partial_{\sigma} \varphi \,\dd\sigma=0,
\label{psigma}
\eeq
The equations of motion following from the above mentioned action permit
a solution exactly describing the folded Frolov-Tseytlin string dual. For this
solution one has
\beq
{\theta^\prime}^{2}-\frac{\omega}{{2\lambda}}  \, \left (
\cos 2 \, \theta -\cos 2 \, \theta_{0}\right )
=0,\quad\quad\varphi=\omega\,t, \label{sigma}
\eeq
which in particular is seen to fulfill the relation~(\ref{psigma}).
The angle $\theta$ can be expressed in terms of the
Jacobi sn function
\beq
\sin\theta(\sigma)=
\sin\theta_0\ \hbox{sn}\!\left.\left (J \sqrt{\frac{ \omega}{{\lambda}}} \,
\sigma \right|\sin^{2} \theta_0\right ),
\label{theta}
\eeq
where the following relation between $\theta_0$ and $\om$ must hold for the
string to be closed and folded exactly once
\beq
J\sqrt{\frac{\om}{\lambda}}=4K(m),
\hspace{0.7cm}m=\sin^2(\theta_0).
\eeq
The angular variable $\theta(\sigma)$ obviously varies in the interval
$[-\theta_0,\theta_0]$..
For any given $\theta_0$ one has (or can impose) the
following identifications, see fig.~\ref{fig:thetaidentifications}.
\beq
\forall n\in\mathds{N},\ \forall z\in\mathds{R},\quad
\ \theta(z+n)=\theta(z)\,,\quad \theta(\ft12-z)=\theta(z)\,.
\label{thetaidentities}
\eeq

In this formulation the
one-loop anomalous
dimension of the gauge theory operator is given by~\cite{Kruczenski:2003gt}
\beq
{\cal E}_0=
\int_{-\ft12}^{\ft12}\theta'(\sigma)^2\,\dd\sigma,
\eeq
and
\beq
\frac{J_2}{J}=\int_{-\ft12}^{\ft12}
\sin^2\theta(\sigma)\,\dd\sigma.
\eeq
which are easily seen to reproduce eqns.~(\ref{E1}) and~(\ref{J2overJ}).

\subsection{Coherent state strings}

The coherent state vectors $|{\bf n}\rangle$  single
out the endpoint of the folded string --- a
property which is not natural from the dual gauge theory perspective as
the dual operator must be cyclically symmetric.\footnote{As mentioned
above, in
the coherent state framework cyclicity
manifests itself via the equation~(\ref{psigma}).}
This, in particular, becomes an issue when we wish to calculate matrix
elements between multi-cut states, cf.\ section~\ref{matrixelements}.

We can ensure cyclicity of the state by averaging over cyclic
translations:
\beq
| {\bf n}\rrangle =\sum_{k=1}^{_n} \prod_{i=1}^{L_m} |\VEC{n_{i+k}}\rangle
\eeq
These averaged states, properly normalized, will now represent our
string states. The inner product is defined as follows.
Given two vectors $|{\bf n}\rangle=\prod_{i=1,L_n}|\VEC{n_i}\rangle$
and $|{\bf m}\rangle=\prod_{j=1,L_m}|\VEC{m_j}\rangle$ one has
\beq
\langle {\bf m} | {\bf n}\rangle= \delta_{L_m,L_n}\prod_{i=1}^{L_m} \langle
\VEC{m_{i}}\,|\,\VEC{n_i}\rangle~, \label{innerproduct}
\eeq
from which the definition of $\llangle m|n\rrangle$ follows.

\section{Matrix Elements of $H_{NP}$.\label{matrixelements}}

With our new states we have
\beq
\frac{\lambda}{J}{\cal E}_0=\frac{\llangle {\bf n}|H_{P}|{\bf n}\rrangle}
{\llangle{\bf n}\,|\,{\bf n}\rrangle}.
\label{energyaverage}
\eeq
We would now like to calculate matrix elements of the one-loop non-planar
dilatation operator between coherent state vectors representing folded
Frolov-Tseytlin strings. It is obvious that acting on a coherent state
vector $|{\bf n}\rangle$ with $H_{NP}$ gives rise to a splitting
of a one-string dual into a two-string dual.
Similarly, acting with $H_{NP}$ on a direct
product two coherent state vectors $|{\bf n}\rangle$ and $|{\bf m}\rangle$
can produce a
one-string dual from a two-string dual. In a more traditional gauge theory
language $H_{NP}$ gives rise to trace splitting and trace joining.
The matrix elements of the non-planar dilatation operator contain
information about the genus one correction to the energy of Frolov-Tseytlin
strings. It is obvious, however, that if we would try to determine
this energy correction by considering $H_{NP}$ a perturbation of $H_P$ we
would have to make use of degenerate perturbation theory. For instance,
if we start from a coherent state vector $|{\bf n}\rangle$
of energy ${\cal E}_0$
as given by~eqn.~(\ref{energyaverage}), cut it vertically once and close
the open ends we obtain another state which up to $1/J$ corrections
is an eigenstate with the same energy.
The same is true if we make $l$ vertical cuts where
$l\ll J$, see figure~\ref{fig:cutstate}. We could also cut with some, not too
large skewness, and still obtain a degenerate state. However, we will
restrict ourselves to straight cut states since in the continuum limit
small skewness should not matter and large skewness takes us out of the
sub-space of degenerate states.
We notice that since $\varphi=\om t$ is
constant along the string the inner product between two coherent states
reduces to
\beq
\langle \VEC{n}_1|\VEC{n}_2\rangle =\cos(\theta_1-\theta_2),
\eeq
which implies that we do not need to worry about $\varphi_i$ at all and
can consistently set $\varphi_i=0$.

\subsection{Normalization of states}

\begin{figure}[htbp]
\centerline{\hbox{
   \epsfxsize=6.0in
   \epsffile{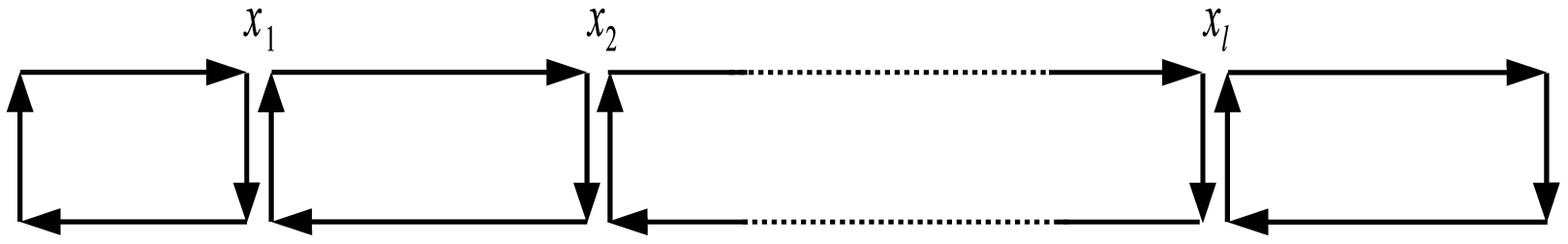}}
  }
  \caption{A cut state $\ket{x_1,\cdots,x_l}$.}\label{fig:cutstate}
\end{figure}

Let us denote by $\ket{\emptyset}$ the complete (uncut) folded string dual, i.e.\
\beq
\ket{\emptyset}\equiv\prod_{i=-J/2}^{J/2}\ket{\VEC{n}_i}~.\label{fullsc}
\eeq
with
\beq
\ket{\VEC{n}_i}=\cos\theta(\ft iJ)\, |\uparrow\rangle +
\sin\theta(\ft iJ)\, |\downarrow\rangle, \quad\quad\quad\quad
-\frac{J}{2}<
i<\frac{J}{2}\,,
\eeq
and with $\theta(x)$ the function given in equation~(\ref{theta}).
Furthermore, let us denote by
$\ket{x_1,\ldots,x_l}$ the state obtained from~(\ref{fullsc}) by cutting
it vertically at the points $x_1,x_2,\ldots x_l$, (see Figure
\ref{fig:cutstate})
\begin{eqnarray}
\lefteqn{
\ket{x_1,\cdots,x_l}\equiv} \nonumber\\
&&\left|\left.\prod_{i=-J/4}^{x_1\,J}
\VEC{n}_{i}\prod_{i=-J/4}^{x_1\,J}  \VEC{n}_{(x_1-\ft14)J-i}\right.
\right\rangle \otimes
\left|\left.\prod_{k=x_1\,J+1}^{x_2\,J}\VEC{n}_i\prod_{k=x_1\,J+1}^{x_2\,J}
\VEC{n}_{(x_1+x_2)\,J+1-i}\right.\right \rangle \otimes
\nonumber\\[3mm]
&&\cdots\otimes\left|\left.\prod_{k=x_l\,J+1}^{J/4}\VEC{n}_i
\prod_{k=x_l\,J+1}^{J/4}
\VEC{n}_{(x_l+\ft14)\,J+1-i}\right.\right\rangle, \label{multitrace}
\end{eqnarray}
where
\beq
-\frac 14<
x_i<\frac14\,, \hspace{1.0cm} l \ll J,\hspace{1.0cm}
x_{j+1}-x_j\sim {\cal O}(J). \label{assumption}
\eeq

\noindent
In order to determine the norm of such a state, we first
consider a single piece of string, extending between the points $x$ and
$y$ and compute
the inner product $\langle \,|\,\rangle$
between this piece and the piece which appears from it by shifting each
of its coherent state vectors a distance $\delta$.
\beq
{\cal A}_{x,y,\delta}\equiv
\langle\prod_{i=x\,J}^{y\,J}
\VEC{n}_{i-\delta\,J}\,|\,
\prod_{i=x\,J}^{y\,J}\VEC{n}_{i}\rangle.
=\prod_{i=0}^{(y-x)\,J} \langle
\VEC{n}_{(x-\delta)\,J+i}\,|\,\VEC{n}_{x\,J+i}\rangle\label{shift-overlap}.
\eeq
For fixed $\delta$, it is clear that ${\cal
A}_{x,y,\delta}$ goes exponentially to zero as $J$ goes to
infinity.
It is therefore sufficient to study
the behavior of ${\cal A}_{x,y,\delta}$ for small $\delta$~: \bea
{\cal A}_{x,y,\delta}&\approx&\exp\left[J \,\int_{x}^{y}
\log\left[\cos\left[\theta(z-\delta)-\theta(z)\right]\right]\,\dd z\right]\nnb\\
&\approx&\exp\left[-J\ft{\delta^2}{2} \,\int_{x}^{y}
\theta'(z)^2\,\dd z\right]\nnb\\
&\approx&\exp\left[-J\ft{\delta^2}{2} \,{\cal
E}_{x,y}\right],\label{overlapdelta}
 \eea
where ${\cal E}_{x,y}$ is
given by
\bea
{\cal E}_{x,y}&\equiv&\int_{x}^{y} \theta'(z)^2\,\dd z \\
&=& 4 \, K(m) \, \bigg(\hbox{E}\left[
\hbox{am}\left ( 4 \, K \, y|m\right ) \right] -\hbox{E}\left[
\hbox{am}\left ( 4 \, K \, x|m\right )\right]- 4
\, K(m)\, \left ( 1-m\right ) \, \left ( y-x\right )  \bigg ).\label{Exy}
\nonumber
\eea
Notice that the planar energy of the folded string stretching between
$x$ and $y$ is $2{\cal E}_{x,y}$ and in particular by
definition ${\cal E}_0={\cal E}_{-\ft12,\ft12}$.
It is then easy to find the square of the
norm of the string with no cuts at leading order in $J$
by integrating over all possible\footnote
{Since we assume that $x_{j+1}-x_j\sim {\cal O}(J)$ ,
 the integration range of such a Gaussian integral can always be taken to be
 $]\!-\infty,+\infty\,[$ when $J\rightarrow \infty$.}
 $\delta$ :
\beq
\llangle\,\emptyset|\emptyset\,\rrangle=J^2\,\int_{-\ft12}^{\ft12}\exp
\left[-J\ft{{\cal E}_0}{2}\,\delta^2\right]\dd \delta=J\sqrt{ \frac{2\,\pi\,J}{{\cal E}_0}}~.\label{norm}
\eeq
One of the factors of $J$ comes from the fact that one can
simultaneously make the same cyclic translation of the bra and the ket
without changing anything. The second factor of $J$ comes from
the summation over nontrivial {\em relative} translations, and the
substitution of a continuous integral for the discrete sum in the large
$J$ limit.
For each smaller string in (\ref{multitrace}), one will
get a similar factor
 so that
\beq
\llangle x_1,x_2,\cdots,x_l|
x_1,x_2,\cdots,x_l\rrangle=
\prod_{i=0}^l \frac{l_i\,J \left(\pi\,J\right)^{\ft {1}{2}}}
{\sqrt{{\cal E}_{x_i,x_{i+1}}}},
\eeq
where
$x_0\equiv-\ft14$,  $x_{l+1}\equiv\ft14$ and $l_i\equiv 2\,(x_{i+1}-x_i)$.

 Here, we neglected the
contributions coming from the ``corners'' of the string pieces
where the overlap is not anymore between $\theta(z-\delta)$
and $\theta(z)$ as in (\ref{overlapdelta}). This is justified
because the relevant shifts $\delta\,J$ are much  smaller than
the length of the pieces we consider.

\subsection{Matrix elements for string joining \label{joining}}

We compute in
this section  the matrix element $\llangle \emptyset|H_{\rm
NP}|x\rrangle$.
To begin with we consider non-cyclic states.

\begin{figure}[h]
\begin{center}
$\begin{array}{cc} \multicolumn{1}{l}{\mbox{$\ket{a}$}} &
   \multicolumn{1}{l}{\mbox{$\ket{b}$}} \\ [-0.2cm]
\epsfxsize=3in \epsffile{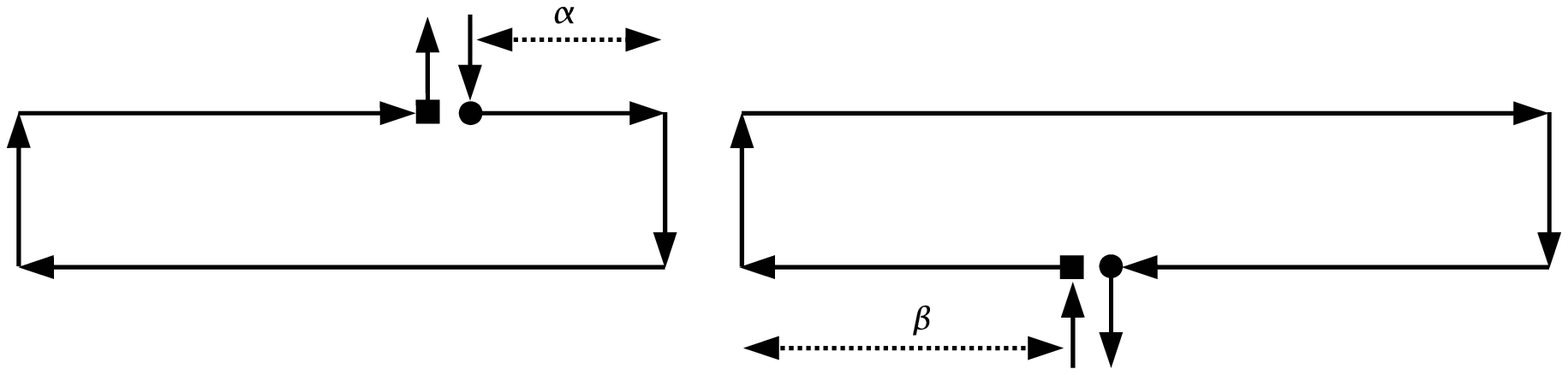} &
    \epsfxsize=3in
    \epsffile{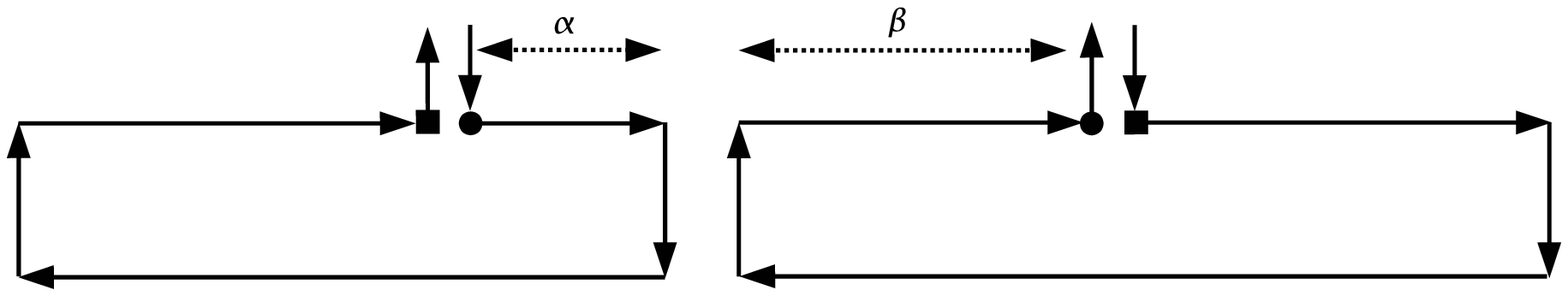} \\
\multicolumn{1}{l}{\mbox{$\ket{c}$}} &
   \multicolumn{1}{l}{\mbox{$\ket{d}$}} \\ [-0.2cm]
\epsfxsize=3in \epsffile{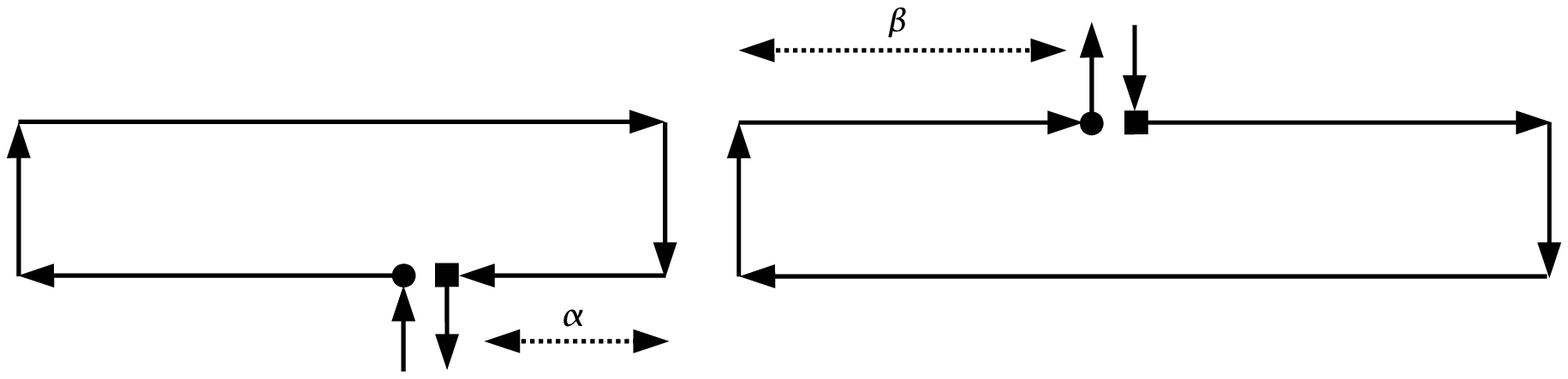} &
    \epsfxsize=3in
    \epsffile{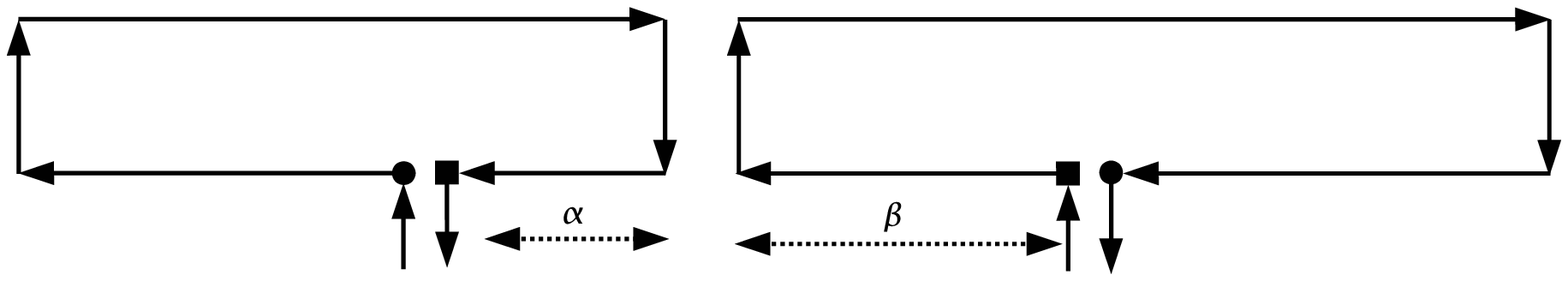} \\
\end{array}$
\end{center}
\caption{Possible joinings of two bits. Sites at the squares
(circles) are linked after  joining and then antisymmetrized.}
\label{figure:joining}
\end{figure}

\begin{figure}[h]
\begin{center}
\epsfxsize=5in \epsffile{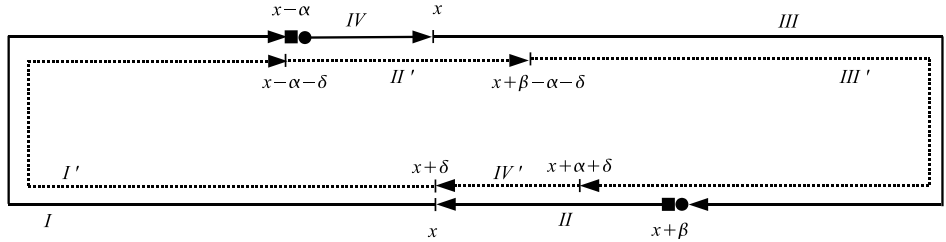}
\end{center}
\caption{Overlaps between the bra $\bra{\emptyset}$ (dotted
lines) and the ket $\ket{a}$ (continuous lines). Arguments for
$\theta(x)$ are given at the relevant points. More precisely,
the bra reads $\bra{I'\ II'\ III'\ IV'}$ and the ket
$\ket{I\ II\ III\ IV}$~: in this figure, it is the function
$\theta(x)$ which is continuous along the loop while the sequence
inside the ket is discontinuous.}
\label{figure:overlap_join}
\end{figure}

There are in total four ways to join a two-piece
state, giving rise to the four different states $\ket{a}$,
$\ket{b}$, $\ket{c}$ and $\ket{d}$ as shown in Figure
\ref{figure:joining}. The eventual use of the cyclic states $| \, \rrangle$
is essential here since the notion of the endpoint of the string becomes
ambiguous.
By reflection symmetry, states $\ket{a}$ and
$\ket{c}$ give the same expectation values, and so do states
$\ket{b}$ and $\ket{d}$. We will start with state $\ket{a}$. The
corresponding overlaps are shown in Figure
\ref{figure:overlap_join}.

As in the previous section, we denote by $\delta$ the
shift given to $\langle\emptyset|$ and by $\bra{I'_\delta}$,
$\bra{II'_\delta}$, $\bra{III'_\delta}$, $\bra{IV'_\delta}$ its
corresponding $\delta$- shifted pieces (see Figure
\ref{figure:overlap_join}). We also define the planar energies of
the first and second spin chain bits respectively by
$$ {\cal E}_1\equiv{\cal E}_{-\ft12-x,x}=2\,{\cal E}_{-\ft14,x}\quad{\rm and }
\quad {\cal E}_2\equiv{\cal E}_{x,\ft12-x}=2\,{\cal E}_{x,\ft14}~.$$
The identity ${\cal E}_0={\cal E}_1+{\cal E}_2$ is satisfied by
construction. First, let us assume that
$\beta\ge\alpha$. We have
$$\llangle \emptyset|a\rrangle=\sum_\delta{\cal F}_{\alpha,\beta,\delta}\,\langle I'_\delta | I \rangle\,
\langle II'_\delta | II \rangle\, \langle III'_\delta | III
\rangle\, \langle IV'_\delta | IV \rangle~,$$
where anti-symmetrization effects at the joining sites are taken into
account through the ${\cal F}_{\alpha,\beta,\delta}$ factor.

~~
\noindent
In order to do the computation, we expand as follows
\beq
\log\left[\cos\left[\theta(z-\eps)-\theta(z)\right]\right]
=-\frac{\eps^2}{2} \,
 \theta'(z)^2+\frac{\eps^3}{2} \,
\theta'(z)\theta''(z)+{\cal O}(\eps^4), \label{logcos}
\eeq
and make use of the identities~(\ref{thetaidentities}) for $\theta(z)$.
It is important to stress
that the expansion we will use for the integrands strongly depends
on the range of integration. For long range integrations,
\emph{e.g} $\int_{-\ft12-x}^{x}f(z,x, \alpha,\beta,\delta)\dd z$,
we expand for small $\alpha,\beta,\delta$'s only. For short range
integrations, \emph{e.g} $\int_{0}^{\beta}f(z,x,
\alpha,\beta,\delta)\dd z$, we also expand for small $z$'s.

One then gets \bea \langle I'_\delta | I \rangle&=&\langle
\prod_{i=(-\ft12-x)\,J}^{(x-\alpha)\,J}
\VEC{n}_{i-\delta\,J}\ ,
\prod_{i=(-\ft12-x)\,J}^{(x-\alpha)\,J}
\VEC{n}_{i}\rangle\nnb\\
 &\approx&\exp\left[J\,\int_{-\ft12-x}^{x-\alpha}
 \log\left[\cos\left[\theta(z-\delta)-\theta(z)\right]\right]\,\dd z\right]\nnb\\
 &\approx&\exp\left[J\,\int_{-\ft12-x}^{x}\left(-\frac{\delta^2}{2} \,
 \theta'(z)^2+\frac{\delta^3}{2} \,
 \theta'(z)\theta''(z)\right)\dd z+J\frac{\delta^2\alpha}{2}\theta'(x)^2\right]\nnb\\
 &\approx&\exp\left[-\frac{1}{2} \, {\cal E}_1 \, J \, \delta ^{2}
 +J\frac{\delta^2\alpha}{2}\theta'(x)^2\right]~,\label{I_I}
 \eea
\bea \langle II'_\delta | II \rangle&=&\langle
\prod_{i=0}^{\beta\,J} \VEC{n}_{(x+\beta-\alpha
-\delta)\,J-i}\ , \prod_{i=0}^{\beta\,J}
 \VEC{n}_{x\,J+i}]\rangle\nnb\\
 &\approx&\exp\left[-\frac{1}{2} \,
\theta'( x)^{2}\,J\,\int_{0}^{\beta}(2\,z+\alpha-\beta+\delta)^2\dd z\right]\nnb\\
 &\approx&\exp\left[
-\frac{1}{6} \, J \, \beta  \, \left ( \beta ^{2}+3 \, \left (
\alpha +\delta \right ) ^{2}\right )  \, \theta ^{\prime }\left (
x\right ) ^{2}
 \right]~,\label{II_II}
 \eea
\bea \langle III'_\delta | III \rangle&=&\langle
\prod_{i=x\,J}^{(\ft12-x-\beta)\,J} \VEC{n}_{(\beta-\alpha
-\delta)\,J+i}\ , \prod_{i=x\,J}^{(\ft12-x-\beta)\,J}
 \VEC{n}_{i}\rangle\nnb\\
 &\approx&\exp\left[J\,\int_x^{\ft12-x}\left(-\ft{( \beta -\alpha -\delta)^2}{2} \,
 \theta'(z)^2+\ft{( \beta -\alpha -\delta)^3}{2} \,
 \theta'(z)\theta''(z)\right)\dd z.\right. \\
&&\left.
\hspace{0.7cm}
+J\ft{( \beta -\alpha -\delta)^2\beta}{2}\theta'(\ft12-x)^2\right]\nnb\\
 &\approx&\exp\left[-\frac{1}{2} \, J \, \left ( \beta -\alpha -\delta \right ) ^{2} \,
 {\cal E}_2+J\,\ft{(\beta -\alpha -\delta)^2\,\beta}{2}\,\theta'(x)^2
 \right]~,\label{III_III}
 \eea
\bea \langle IV'_\delta | IV \rangle&=&\langle
\prod_{i=0}^{\alpha\,J} \VEC{n}_{(x+\delta)\,J+i },
\prod_{i=0}^{\alpha\,J}
 \VEC{n}_{x\,J-i}\rangle\nnb\\
 &\approx&\exp\left[-\frac{1}{2} \,
\theta'( x)^{2}\,J\,\int_{0}^{\alpha}
 \left ( 2 \, z+\delta \right ) ^{2} \dd z\right]\nnb\\
 &\approx&\exp\left[-\frac{1}{6} \, J \, \alpha  \, \left ( 4 \, \alpha ^{2}+6 \, \alpha  \, \delta +3 \, \delta ^{2}\right )  \, \theta ^{\prime }\left ( x\right ) ^{2}
 \right]~.\label{IV_IV}
 \eea
The four overlaps in total give the contribution
$$\begin{array}{l}
\exp\left[-\frac{1}{2} \, {\cal E}_1 \, J \, \delta ^{2}
-\frac{1}{2} \, {\cal E}_2 \, J \, \left ( \beta -\alpha -\delta \right ) ^{2}\right.\\
\quad\quad -\frac{1}{3} \,J\, \theta'(x)^2\, \left ( -\alpha ^{3}+3
\, \alpha ^{2} \, \beta +2 \, \beta ^{3} -3 \, \left ( \beta
^{2}+\alpha ^{2}\right )  \, \left ( \beta -\alpha -\delta \right
) \right )\bigg],
\end{array}$$
and one can see that the dominant region will be around
$\delta\approx 0$ and $\beta\approx \alpha$, so that the leading
term in $\ft1J$ will be given by taking the following
approximation for the exponential~:
$$\exp\left[-\frac{1}{2} \, {\cal E}_1 \, J \, \delta ^{2}
-\frac{1}{2} \, {\cal E}_2 \, J \, \left ( \beta -\alpha\right )
^{2}\ -\frac{4}{3} \, J\,\alpha ^{3}\, \theta'(x)^2\right].$$
We should now compute ${\cal F}_{\alpha,\beta,\delta}$ near these
values of  $\alpha$, $\beta$ and $\delta$. One gets\footnote{We
use the notation $f(\rnode{rA}A,\rnode{rB}B) \,g(\rnode{rC}C,\rnode{rD}D)
\ncbar[linewidth=.01,nodesep=2pt,arm=.1,angle=-90]{-}{rA}{rB}
\ncbar[linewidth=.01,nodesep=2pt,arm=.1,angle=-90]{-}{rC}{rD}
=\left(f(A,B)-f(B,A)\right)\,g(C,D)+f(A,B)\,\left(g(C,D)-g(D,C)\right)$.}

\bea
{\cal F}_{\alpha,\alpha,0}&=& \, \left < \VEC{n}_{(x-\alpha-\delta)\,J
}\,,\rnode{AA}{\VEC{n}_{(x-\alpha)\,J  } }\right
>
  \, \left < \VEC{n}_{(x-\alpha-\delta)\,J +1}\,,\rnode{BB}{\VEC{n}_{(x+\beta)\,J }   }\right >\nnb\\[5mm]
&&\quad\quad\quad\quad
 \times\,  \left < \VEC{n}_{(x+\alpha+\delta)\,J +1 }\,,
\rnode{CC}{\VEC{n}_{(x+\beta)\,J+1 }}\right >\,
  \left < \VEC{n}_{(x+\alpha+\delta)\,J}\,,
\rnode{DD}{\VEC{n}_{(x-\alpha)\,J+1}}\right >
\ncbar[linewidth=.01,nodesep=2pt,arm=.3,angle=-90]{-}{AA}{BB}
\ncbar[linewidth=.01,nodesep=2pt,arm=.3,angle=-90]{-}{CC}{DD}
\Bigg|_{\!\!\!\scriptsize
     \begin{array}{l}\delta=0\\ \beta=\alpha\end{array}}\nnb \\[5mm]
&\approx&\frac{4 }{J} \alpha \, \theta'(x)^2~.
\eea
The case
$\alpha>\beta$ gives the same result up to the exchange
$\alpha\leftrightarrow\beta$. Furthermore, translating the result to cyclic
states implies multiplying by $l_1\,l_2\,J^2$.
Finally,
using the normalization factor
${\cal N}=\left(\sqrt{\frac{2\,\pi\,J}{{\cal E}_0}}\,l_0\,J\right)^\ft12\,
\left(\sqrt{\frac{2\,\pi\,J}{{\cal E}_1}}\,l_1\,J\right)^\ft12\,
\left(\sqrt{\frac{2\,\pi\,J}{{\cal E}_2}}\,l_2\,J\right)^\ft12\,$,
one then gets at leading order in
$\ft1J$~:
\bea \sum_{\alpha,\beta}\llangle
\emptyset|a\rrangle&\approx&\frac2{\cal N}\, \frac{4}{J}\, \,\theta'(x)^2
 J^{5}\, l_1\, l_2 \, \int_{0}^{\infty }\!\!\dd\beta\int_{0}^{\beta
}\!\!\dd\alpha \int_{-\infty }^{\infty }\!\!\dd\delta\ \
  e^{-\frac{1}{2} \, {\cal E}_1 \, J \, \delta ^{2}
-\frac{1}{2} \, {\cal E}_2 \, J \, \left ( \beta -\alpha\right )
^{2}\
-\frac{4}{3} \, J \, \alpha ^{3}\, \theta'(x)^2}\,\alpha\nnb\\[3mm]
&\approx&\frac{4  \, \Gamma\left ( \frac{2}{3}\right
)}{3^{1/3} }\, K^{2/3} \, m^{1/3}   \, \hbox{cn}\left ( 4 \, K \,
x|m\right ) ^{2/3} \,
\left(\frac{l_1\,l_2}{l_0}\right)^{1/2}
\left(\frac{2\,\pi\,{\cal E}_0}
{{\cal E}_1\,{\cal E}_2}\right)^{\ft14}\,J^{1/12}~.
\eea
 Note that although $\beta$ should be in
the interval $[0,\ft14-x]$ and $\delta$ in the interval
$[-\ft12,\ft12]$, integrating in both cases till infinity will not
change the leading $\ft1J$ behavior as the integrand converges
exponentially  to zero for $\alpha\,J^{1/3}\gg1$,
$\beta\,J^{1/3}\gg1$ and $\delta\,J^{1/2}\gg1$.

A similar computation shows that $\llangle
\emptyset|b\rrangle$ and $\llangle
\emptyset|d\rrangle$ are of
order $J^{-1/4}$ and therefore can be neglected compared to the
$J^{1/12}$ behavior found here. Thus, one obtains at leading order in
$\ft1J$
\beq
\llangle \emptyset|H_{\rm NP}|x\rrangle= \frac{8  \, \Gamma\left ( \frac{2}{3}\right
)}{3^{1/3} }\, K^{2/3} \, m^{1/3}   \, \hbox{cn}\left ( 4 \, K \,
x|m\right ) ^{2/3} \,
\left(\frac{l_1\,l_2}{l_0}\right)^{1/2}
\left(\frac{2\,\pi\,{\cal E}_0}
{{\cal E}_1\,{\cal E}_2}\right)^{\ft14}\,J^{1/12}~.
\eeq
It is straightforward to generalize this result
to
an arbitrary
number of cuts where the joining takes place at position $x_i$.
It is in order to facilitate this generalization that we have explicitly
kept the parameter $l_0$ although in our case we have $l_0=1$.
We observe the occurrence of the factor $({\cal E}_1\,{\cal E}_2)^{-1/4}$ which diverges when $x$
approaches the endpoints of the string. In this situation we can thus
not trust the semi-classical analysis (and hence the overall $J$-scaling).

\subsection{Matrix elements for string splitting \label{splitting}}
>From the calculations in the last section,
we learn which approximations we are allowed to do in order
to keep only the leading order in $\ft1J$. First, the terms that
arise from the cyclicity of the traces are long range terms : they
appear through $\delta$-shifts over a whole piece of spin chain
and consequently will give in the exponential a square term
times minus the planar energy of the considered piece,
times $J$. This is what happened in equations (\ref{I_I}) and
(\ref{III_III}). Conversely, terms which are integrated on short
intervals will appear in the exponential starting at the cubic
order (see equations (\ref{II_II}) and (\ref{IV_IV})). This allows
for the following approximations that will not change the leading
$\ft1J$ term after all integrations~:
\begin{enumerate}
\item When computing overlaps over long range parts, it is not
necessary to take into account small parameters at the endpoints
of the integration. For example, taking $\int_{-\ft12-x}^{x}\dd z$
instead of $\int_{-\ft12-x}^{x-\alpha}\dd z$ in (\ref{I_I}) would
not have changed the final result. \item When computing
overlaps over short range parts, one can do as if the shifts
appearing in the long range terms were equal to zero.
\end{enumerate}

\begin{figure}[h]
\centerline{\hbox{
   \epsfxsize=5.0in
   \epsffile{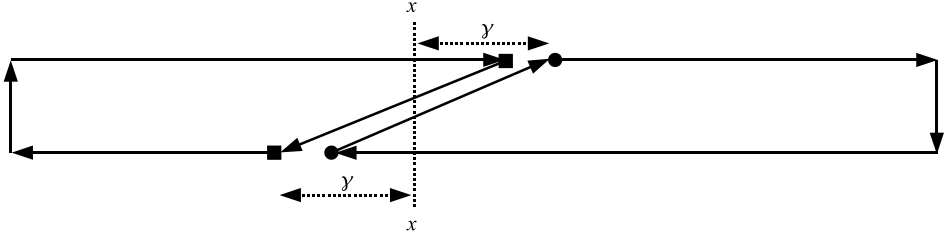}}
  }
  \caption{A state $\ket{\{x,\gamma\}}$. Sites at the squares are antisymmetrised, as sites at the circles. The
spin chain was cut between sites where $\theta$ takes the value
$\theta(x+\gamma)$ and
$\theta(x-\gamma)$.
  }\label{figure:cutting}
\end{figure}

We can now compute expectation values such as $\llangle x|H_{\rm
NP}| \emptyset\rrangle$. $H_{\rm NP}\ket{\emptyset}$ will give a lot of
possible double-chain states. Only the ones with lengths equal to
those of $\ket{x}$, \emph{i.e.} states with length
$(\ft12+2\,x)\,J$ and length $(\ft12-2\,x)\,J$, will contribute.
All these contributing
states can be characterized by a value $\gamma\, J$ expressing
how far the cut took place from the straight cut between sites
$x\,J$ and sites $(\ft12-\,x)\,J$ (see Figure
\ref{figure:cutting}). Let us denote them $\ket{\{x,\gamma\}}$. The
following identity holds~:
$$\llangle x|H_{\rm NP}| \emptyset\rrangle=\sum_{i=-\ft J2}^{\ft J2}\llangle x|\{x,\ft iJ\}\rrangle~.$$
\begin{figure}[h]
\centerline{\hbox{
   \epsfxsize=5.0in
   \epsffile{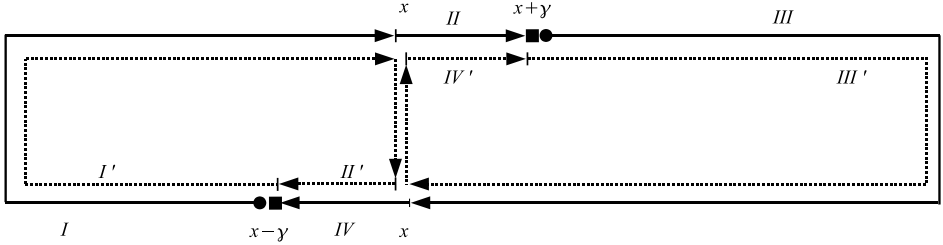}}
  }
  \caption{Overlaps between $\bra{x}$ (dotted lines) and $\ket{\{x,\gamma\}}$ (continuous lines). Arguments for
$\theta(x)$ are given at the relevant points. More precisely,
$\bra{x}$ reads $\bra{I'\ II'}\,\bra{\ III'\ IV'}$
while $\ket{\{x,\gamma\}}$ is equal to $\ket{I\
II}\,\ket{ III\ IV}$.
As in Figure \ref{figure:overlap_join}, it is the function
$\theta(x)$ which is continuous along the loop. Possible shifts
$\delta$ and $\delta'$ for each piece of $\bra{x}$ were put to $0$
for simplicity.
  }\label{figure:overlapcut}
\end{figure}
Overlaps  for $\langle x|\{x,\ft iJ\}\rangle$ are shown in
Figure \ref{figure:overlapcut}. In order to go to the full cyclic
scalar product, one should then add two arbitrary shift $\delta$
and $\delta'$ for each piece of $\bra{x}$ as well as one for
$|\emptyset\rangle$. However, the effect of the latter is simply
the multiplication by the factor $l_0\,J$.

We thus have
$$\llangle x|H_{\rm NP}|\emptyset\rrangle=l_0 J\,\sum_{\gamma,\delta,\delta'}{\cal F}_{\gamma,\delta,\delta'}\,
\langle I'_\delta | I \rangle\, \langle II'_\delta | II \rangle\,
\langle III'_{\delta'} | III \rangle\, \langle IV'_{\delta'} | IV
\rangle~,$$
where ${\cal F}_{\gamma,\delta,\delta}$ is the
anti-symmetrization factor and $\bra{I'_\delta}$,
$\bra{II'_\delta}$, $\bra{III'_{\delta'}}$, $\bra{IV'_{\delta'}}$
are the $\delta$ ($\delta'$) shifted pieces of $\bra{x}$.

Using the approximations we presented at the beginning of this
section, we have, for $\gamma>0$,
\bea
\langle I'_\delta | I \rangle &\approx&
\langle \prod_{i=(-\ft12-x)\,J}^{x\,J} \VEC{n}_{i-\delta\,J}\
, \prod_{i=(-\ft12-x)\,J}^{\,J}
 \VEC{n}_{i}\rangle\nnb\\
 &\approx&\exp\left[-\frac{1}{2} \, {\cal E}_1 \, J \, \delta ^{2}
\right]~,\label{I_Icut}
 \eea
\bea \langle II'_\delta | II \rangle&\approx&\langle
\prod_{i=0}^{\gamma\,J} \VEC{n}_{x \,J-i}\ ,
\prod_{i=0}^{\gamma\,J}
 \VEC{n}_{x\,J+i}\rangle\nnb\\
 &\approx&\exp\left[-2 \,
\theta'( x)^{2}\,J\,\int_{0}^{\gamma}z^2\dd z\right]\nnb\\
 &\approx&\exp\left[
-\frac{2}{3} \, J \, \gamma^3 \, \theta ^{\prime }\left ( x\right
) ^{2}
 \right]~,\label{II_IIcut}
 \eea
\bea \langle III'_{\delta'} | III \rangle&\approx&\langle
\prod_{i=x\,J}^{(\ft12-x)\,J} \VEC{n}_{i- \delta'\,J}\ ,
\prod_{i=x\,J}^{(\ft12-x)\,J}
 \VEC{n}_{i}\rangle\nnb\\
 &\approx&\exp\left[-\frac{1}{2} \, J \, {\delta'} ^{2} \,
 {\cal E}_2
 \right]~,\label{III_IIIcut}
 \eea
\bea \langle IV'_\delta | IV \rangle&\approx&\langle
\prod_{i=0}^{\gamma\,J} \VEC{n}_{x \,J+i}\ ,
\prod_{i=0}^{\gamma\,J}
 \VEC{n}_{x\,J-i}\rangle\nnb\\
 &\approx&\exp\left[-2 \,
\theta'( x)^{2}\,J\,\int_{0}^{\gamma}z^2\dd z\right]\nnb\\
 &\approx&\exp\left[
-\frac{2}{3} \, J \, \gamma^3 \, \theta ^{\prime }\left ( x\right
) ^{2}
 \right]~.\label{IV_IVcut}
\eea The overlaps therefore give the contribution
$$\begin{array}{l}
\exp\left[-\frac{1}{2} \, {\cal E}_1 \, J \, \delta ^{2}
-\frac{1}{2} \, {\cal E}_2 \, J \, {\delta'} ^{2} -\frac{4}{3} \,
\theta'(x)^2\,J\, \gamma ^{3}\right]~.
\end{array}$$
Computing ${\cal F}_{\gamma,\delta,\delta'}$ around
$\delta=\delta'=\gamma=0$, one gets \bea {\cal F}_{\gamma,0,0}&=&
\, \left < \VEC{n}_{(x-\gamma)\,J+1   }\,,\rnode{AA}{\VEC{n}_{(x+\gamma)\,J  }
}\right >
  \, \left < \VEC{n}_{(x-\gamma)\,J}\,,\rnode{BB}{\VEC{n}_{(x-\gamma)\,J} }\right >\nnb\\[5mm]
&&\quad\quad\quad\quad
 \times\,  \left < \VEC{n}_{(x+\gamma)\,J}\,,\rnode{CC}{\VEC{n}_{(x-\gamma)\,
J+1 }}\right >\,
  \left < \VEC{n}_{(x+\gamma)\,J+1}\,,\rnode{DD}{\VEC{n}_{(x+\gamma)\,J+1}}\right >
\ncbar[linewidth=.01,nodesep=2pt,arm=.3,angle=-90]{-}{AA}{BB}
\ncbar[linewidth=.01,nodesep=2pt,arm=.3,angle=-90]{-}{CC}{DD}
\nnb\\[5mm]
&\approx&\frac{4 }{J} \gamma \, \theta'(x)^2~. \label{vertexfactor}
\eea
In the $\gamma<0$ case, extra minus signs appear so that one can use the same
results by taking the absolute value of $\gamma$ instead.
Using as normalization the factor ${\cal N}=\left(\sqrt{\frac{2\,\pi\,J}{{\cal E}_0}}\,l_0\,J\right)^\ft12\,
\left(\sqrt{\frac{2\,\pi\,J}{{\cal E}_1}}\,l_1\,J\right)^\ft12\,
\left(\sqrt{\frac{2\,\pi\,J}{{\cal E}_2}}\,l_2\,J\right)^\ft12\,$,
this leads to
\bea
&&\hspace{-0.9cm}\llangle x|H_{\rm NP}|
\emptyset\rrangle \approx \frac1
{\cal N}\frac{4}{J}\, \,\theta'(x)^2 \,l_0\,
J^{4}\,  \int_{-\infty}^{\infty }\!\!\dd\gamma\int_{-\infty}^{\infty
}\!\!\dd\delta \int_{-\infty }^{\infty }\!\!\dd\delta'\ \
  e^{-\frac{1}{2} \, {\cal E}_1 \, J \, \delta ^{2}
-\frac{1}{2} \, {\cal E}_2 \, J \, {\delta'} ^{2}\
-\frac{4}{3} \, J \, |\gamma| ^{3}\, \theta'(x)^2}\,|\gamma|\nnb\\[3mm]
&& \approx \frac{8  \, \Gamma\left ( \frac{2}{3}\right
)}{3^{1/3} }\, K^{2/3} \, m^{1/3}   \, \hbox{cn}\left ( 4 \, K \,
x|m\right ) ^{2/3} \,
\left(\frac{l_0}{l_1\,l_2}\right)^{1/2}
\left(\frac{2\,\pi\,{\cal E}_0}
{{\cal E}_1\,{\cal E}_2}\right)^{\ft14}\,J^{-11/12}. \label{rescut}
\eea
This result can be
immediately extended to states which were already cut before
the action of the Hamiltonian.
We note that the non-planar dilatation operator is non-hermitian. A similar
situation was encountered in previous analyses of the non-planar corrections
to energies of BMN states~\cite{Janik:2002bd,Beisert:2002ff}.
There the non-planar
dilatation operator was related to its hermitian conjugate by a similarity
transformation. The same is the case here.

\section{A solvable toy model \label{integrable}}

By construction the vertically cut multi-string states studied above are
degenerate in planar energy with the complete Frolov-Tseytlin string.
Let us now consider a toy
model of a folded string for which the vertically cut states exhaust the
space of states degenerate in energy with the uncut string. Furthermore,
let us assume that the matrix elements of $H_{NP}$
for string splitting and string joining depend only on the point of
splitting and joining. Determining the first non-planar correction to
the string energy under these assumptions amounts to diagonalizing the
non-planar dilatation operator in the subspace of vertically cut states
which of course implies diagonalizing an infinite dimensional matrix in
the limit $J\rightarrow \infty$. This problem can easily be solved, however.
Let us denote by $\ket{i,j,k,\cdots}$ the state corresponding to
the string cut at  positions $i,j,k,\cdots$ and by
${\cal X}_l$ the matrix element corresponding
an additional cut or joining taking place
at position $l$. To illustrate the solution,
we consider as an example only three possible sites where a cut/joining
can take place. Then in the base $\{\ket{\emptyset},\ket{1},\ket{2},\ket{1,2},\ket{3},\ket{1,3},\ket{2,3},
\ket{1,2,3}\}$, the matrix we have to diagonalize is given by
$${\cal M}=\left(
\begin{array}{llllllll}
 0 & {\cal X}_1 & {\cal X}_2 & 0 & {\cal X}_3 & 0 & 0 & 0 \\
{\cal X}_1 & 0 & 0 & {\cal X}_2 & 0 & {\cal X}_3 & 0 & 0 \\
 {\cal X}_2 & 0 & 0 & {\cal X}_1 & 0 & 0 & {\cal X}_3 & 0 \\
 0 & {\cal X}_2 & {\cal X}_1 & 0 & 0 & 0 & 0 & {\cal X}_3 \\
 {\cal X}_3 & 0 & 0 & 0 & 0 & {\cal X}_1 & {\cal X}_2 & 0 \\
 0 & {\cal X}_3 & 0 & 0 & {\cal X}_1 & 0 & 0 & {\cal X}_2 \\
 0 & 0 & {\cal X}_3 & 0 & {\cal X}_2 & 0 & 0 & {\cal X}_1 \\
 0 & 0 & 0 & {\cal X}_3 & 0 & {\cal X}_2 & {\cal X}_1 & 0
\end{array}
\right)$$
whose eigenvalues $\mu$ are simply all possible sum and differences between the ${\cal X}_i$'s :
$$\mu=\pm {\cal X}_1\pm {\cal X}_2\pm {\cal X}_3~.$$
In the case of $J$ different sites, the eigenvalues
are distributed in a
quasi-continuum between energies $\dst\pm J\,\int_{-\ft14}^{\ft14}{\cal X}_x\,\dd x$. In our case we can arrange by means of a similarity transformation
that all our matrix elements scale as $J^{-5/12}$. Therefore, a rough scaling
argument gives
\beq
\Delta E\approx \frac{\lambda}{N}
\frac J2 \,{\cal X}_0 \sim \lambda  \, \, \frac{J^{7/12}}{N}.
\label{unexpected}
\eeq
\noindent
Now if one, again naively, assumes BMN-like scaling for the energy of
spinning strings one needs that the genus
one contribution compared to the genus zero one has an additional factor of
$\frac{J^2}{N}$ which leads to the expectation
$\Delta E \sim \frac{J}{N}$. It is of course not known
to which extent BMN scaling beyond the planar limit
should hold for spinning strings. One knows from the analysis
of~\cite{Eden:2006rx,Beisert:2006ez} and the field theoretical computations
of~\cite{Bern:2006ew} that BMN scaling for few-impurity
operators breaks down already at the planar level but only at
order four in $\lambda$. In the true picture of string splitting we can
not claim that the straight cut states exhaust the space of eigenstates
degenerate in energy with the folded string.\footnote{As mentioned
earlier the straight cut states are also not exact eigenstates but
only eigenstates up to terms of order $\frac{1}{J}$.}
One could argue that one
should in fact replace ${\cal X}_x$ of the toy model by some integral
over matrix elements involving skew cut states close to the vertically
cut ones and that this could give rise to additional factors of $J$.
We have not been able to make a quantitative estimate of this effect,
but we
find it unlikely that such an integration could
provide the ``missing'' factor $J^{5/12}$. Rather the low power of $J$
i eqn.~(\ref{unexpected}) seems to suggest
that the process of semi-classical string
splitting and joining is not of importance for the genus one energy
shift, cf.\ section~\ref{discussion}.

\section{Discussion \label{discussion} }

Our calculation shows that for long strings a nonzero contribution to the
splitting matrix element comes only from strings which are almost on top of
each other, cf.\ eqn.~(\ref{logcos}) and subsequent calculations.
This is somewhat reminiscent of the interaction vertex between
strings in light cone string field theory:
\bea
\lefteqn{V(X^i_0(\sigma),X^i_1(\sigma),X^i_2(\sigma))= } \\
&&\int ds_0 ds_1 ds_2 \, \delta(J_0-J_1-J_2) \times \nonumber \\
&&
\prod \Delta(X^i_1(\sigma+s_1)-X^i_0(\sigma+s_0))
\Delta(X^i_2(\sigma+s_2)-X^i_0(\sigma+s_0+J_1/J_0)). \nonumber
\eea
In the above formula the $s_i$ are direct analogues of cyclic translations in
our definition of states, while the functional delta functions are analogues
of the property that we have found namely that in order for the matrix
element to be nonzero the angles defining the coherent states have to be
within $J^{-1/2}$. However the detailed calculations in
sections~\ref{joining} and~\ref{splitting} show
that more nontrivial $J^{-1/3}$ factors may also appear. In addition we saw
that the $H_{NP}$ operator gives an effective additional operator inserted at
the interaction point, cf.\ eqn~(\ref{vertexfactor}). This is not unexpected since such operators
appear generically in superstring light cone SFT
(see e.g.~\cite{Roiban:2002xr}.)
However, due to the fact that we really can deal only with classical
states we refrain from making any more quantitative comparison.

Our crude estimate of the order of magnitude of the genus one energy
shift due to semi-classical string joining and splitting lead to the energy
scaling  with an unexpectedly small power of $J$.
An interpretation of this result may be that the
contribution to the energy shift coming from such semi-classical string
processes is simply quite small. In fact for generic macroscopic rotating
strings (i.e. not `folded' ones) the contribution of string splitting
into classical states would be very strongly suppressed. It is
much more probable that the dominant non-planar contribution would come
from {\em small} strings which would split off from the rotating
string and which would be reabsorbed shortly after.
Unfortunately the
process of small strings splitting off
is beyond the reach of the semi-classical coherent
state methods which we were using.

\vspace*{1.0cm}

\noindent
{\bf Acknowledgments}
P.-Y. Casteill  thanks F. Morales and C. Sochichiu for enlightening discussions
and comments. C.\ Kristjansen acknowledges the support of
the Banff International Research Station during participation in
the focused research group on
``Integrability, Gauge Fields and Strings'' in July 2007.

The authors were all supported by
ENRAGE (European Network on Random Geometry), a Marie Curie
Research Training Network financed by the European Community's
Sixth Framework Program, network contract MRTN-CT-2004-005616.
RJ was supported in part by Polish Ministry of Science and Information
Society Technologies grant 1P03B04029 (2005-2008). CK
was supported in part by FNU through grant number 272-06-0434

\end{document}